\documentclass[aps,pra,groupedaddress,showpacs,twocolumn,sort&compress,
floatfix, amssymb, noeprint,10pt]{revtex4-2}
\usepackage[english]{babel}
\usepackage[dvips]{graphicx}
\usepackage{dcolumn}

\usepackage[hidelinks]{hyperref}
\usepackage{footnote}
\usepackage{siunitx}
\usepackage{booktabs}
\usepackage{subcaption}

\usepackage{amsmath} 
\usepackage{caption}
\renewcommand{\selectlanguage}[1]{}
\begin{document}

\title{Rapid stochastic  spatial light modulator calibration and pixel crosstalk optimisation}

\author{ P.\ Schroff, E.\ Haller, S.\ Kuhr\ and A.\ La Rooij}
 \email{arthur.larooij@strath.ac.uk}  
 
\affiliation{University of Strathclyde, Department of Physics, SUPA, Glasgow G4 0NG, United Kingdom}

\begin{abstract}
Holographic light potentials generated by phase-modulating liquid-crystal spatial light modulators (SLMs) are widely used in quantum technology applications.
Accurate calibration of the wavefront and intensity profile of the laser beam at the SLM display is key to the high fidelity of holographic potentials.
Here, we present a new calibration technique that is faster than previous methods while maintaining the same level of accuracy.
By employing stochastic optimisation and random speckle intensity patterns, we calibrate a digital twin that accurately models the experimental setup.
This approach allows us to measure the wavefront at the SLM to within $\lambda/170$ in $\sim\!5~\text{minutes}$ using only 10 SLM phase patterns, a significant speedup over state-of-the-art techniques.
Additionally, our digital twin models pixel crosstalk on the liquid-crystal SLM, enabling rapid calibration of model parameters and reducing the error in light potentials by a factor of $\sim\!5$ without losing efficiency.
Our fast calibration technique will simplify the implementation of high-fidelity light potentials in, for example, quantum-gas microscopes and neutral-atom tweezer arrays where high-NA objectives and thermal lensing can deform the wavefront significantly.
Applications in the field of holographic displays that require high image fidelity will benefit from the novel pixel crosstalk calibration, especially for displays with a large field of view and increased SLM diffraction angles.

\end{abstract}

\maketitle

\pagenumbering{arabic}

\section{Introduction}
\label{sec:intro}

Liquid crystal on silicon (LCOS) SLMs have become a useful tool in fields ranging from holographic displays to state-of-the-art quantum computing \cite{browaeys2020many,henriet2020quantum, graham2022multi, wurtz2306aquila, morgado2021quantum, nikolov2023randomized, savage2009digital}.
Specifically, the rapidly advancing field of neutral-atom quantum computing has benefited from highly uniform and efficient light potentials that have been generated holographically using phase-modulating LCOS SLMs \cite{Ebadi2021, Barredo2016, Amico2021, manetsch2024tweezer, Bruce2015}.
Outside the scope of cold-atom experiments, tailored light potentials are used in biomedical applications such as optogenetic stimulation \cite{adesnik2021probing}, non-invasive imaging through tissue \cite{may2021fast}, and high-resolution 3D imaging \cite{Li_Wu2023} and tomography \cite{huang_quantitative_2024}.
Recent advances in holographic displays for virtual and augmented reality applications have been driven by machine learning techniques, greatly improving image quality and reducing the time taken to generate holograms \cite{Choi2021, Peng2020, Jang2024}.
Phase-modulating LCOS SLMs imprint a phase pattern onto the incident laser beam, achieving tailored phase and intensity distributions in the image plane located in the far field or the Fourier plane of a lens.
To determine the SLM phase pattern that creates the desired potential in the image plane, one must solve the so-called phase retrieval problem.
Several iterative phase retrieval algorithms have been developed, achieving simulated light potentials with root-mean-squared (RMS) errors well below $\SI{1}{\%}$ \cite{deMarco2008, Harte2014, Gaunt2012, Peng2020, Choi2021, sui_2024}.
However, these algorithms do not account for experimental effects and assume that the intensity profile of the incident laser beam and the phase of the light immediately after the SLM are perfectly known.

In practice, aberrations introduced by optical elements cause imperfections in the intensity profile and wavefront of the incident laser beam.
The aberrated wavefront and deformations of the SLM's reflective surface cause a spatially varying phase offset in the plane of the SLM which adds to the phase delay introduced by each SLM pixel.
To increase the uniformity of the experimental potentials, both the intensity profile of the incident laser beam and the phase at the SLM must be precisely measured \cite{Zupancic2016, Schroff2023}.
Furthermore, the fringing field effect leads to crosstalk between neighbouring SLM pixels, causing inhomogeneities in the light potential \cite{Moser2019} that phase retrieval algorithms typically do not address.
The magnitude of pixel crosstalk depends on the thickness of the liquid crystal layer and the size of the pixel electrodes - smaller pixel sizes and thicker liquid crystal layers amplify the effect.
As manufacturers steadily increase the resolution of modern LCOS SLMs and consequently shrink their pixel pitch, it has become increasingly important to model pixel crosstalk to compensate for its effects.
Recently, accurate light potentials have been generated by limiting the gradient of the SLM phase pattern, reducing the effect of pixel crosstalk \cite{ropac_2024}.
It has also been shown that by modelling pixel crosstalk using a convolution, experimental errors in the resulting light potentials can be reduced \cite{Schroff2023, Ronzitti2012, Pushkina2020, Jang2024, Buske2024, persson_reducing_2012}.
However, finding the optimal convolution kernel for the SLM phase pattern is time-consuming as it involves measuring the intensity of higher diffraction orders \cite{Moreno2021} or iteratively optimising the camera image of a light potential \cite{Schroff2023, Buske2024}.
Several schemes have been developed to measure the phase and the intensity profile at the SLM.
Methods that measure the phase in a few seconds using a Twyman-Green interferometer have been proposed \cite{Li2023, Lee2023}, however, they require additional optical elements and their accuracy relies on the flatness of a reference mirror.
Self-interfering calibration schemes that do not require such additional components are typically slow as they need to display hundreds of phase patterns on the SLM to achieve the desired spatial resolution and are not straightforward to implement \cite{Zupancic2016, Supikov2023, Zhao2018, Zhao2022, Schroff2023}.

In this work, we present a technique to measure the phase and intensity profile in a Fourier imaging setup that does not require any additional hardware.
Our scheme is faster than previous self-interfering methods \cite{Zupancic2016, Schroff2023, Supikov2023}, requiring significantly fewer camera images and maintaining their high accuracy.
Inspired by recent work \cite{Wu2019, Peng2020}, we calibrate a digital twin of the experimental setup by minimising the difference between random speckle images captured by the camera and those simulated by the digital twin (Fig.~\ref{fig:fast_calib_schemMain}).
In the second part of this study, we use our digital twin to simulate pixel crosstalk and thereby optimise the parameters of different crosstalk models in $\sim\SI{2}{minutes}$.
In addition to accounting for pixel crosstalk using a convolution \cite{Pushkina2020, Moreno2021, Ronzitti2012, Buske2024, Jang2024}, we propose a model that is more efficient to compute, despite containing a larger number of parameters \cite{Moser2019}.
We benchmark the accuracy of the different crosstalk models by generating a top-hat potential using a camera feedback algorithm \cite{Schroff2023} and demonstrate that the optimum model reduces the error of the light potentials using fewer camera feedback iterations \cite{Schroff2023}.

\section{Stochastic optimisation technique}
\label{sec:Stoch}

\begin{figure}[t] \centering{}
    \includegraphics[width = 0.95 \linewidth]{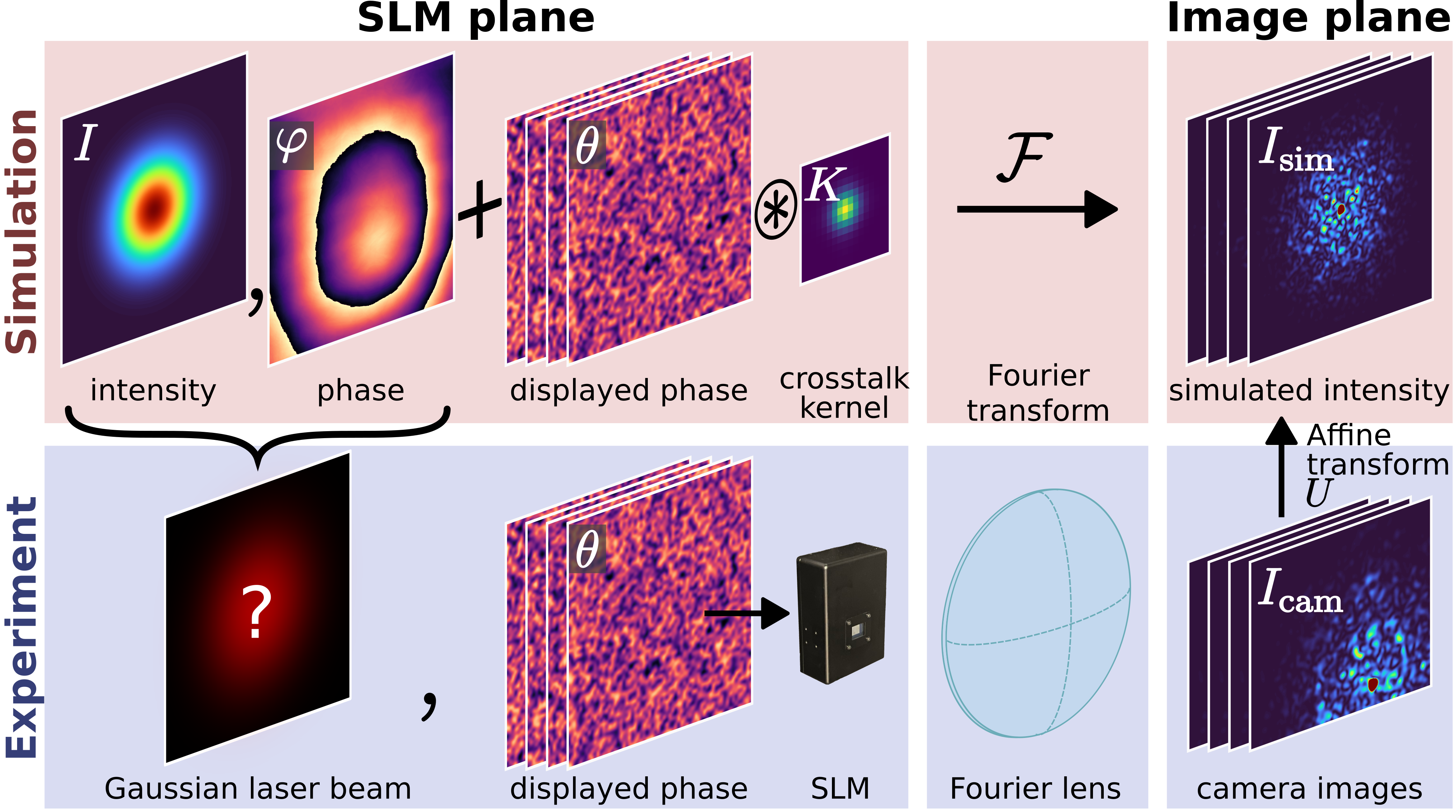}
    \caption{A digital twin (red box) simulates our experimental Fourier imaging setup (blue box). To train the digital twin, parameters of the simulation (the laser beam intensity, $I$, and phase at the SLM, $\varphi$, the pixel crosstalk kernel, $K$, and the affine transformation matrix, $U$) are adjusted to minimise the difference between the simulated images, $I_{\text{sim}}$, and the camera images, $I_{\text{cam}}$, when displaying random phase patterns, $\theta$, on the SLM. The physical lens in the experimental setup is modelled using a Fourier transform.
    }
\label{fig:fast_calib_schemMain}
\end{figure}

Current state-of-the-art techniques measure the phase by utilising the interference from multiple beams diffracted by small SLM phase patterns that are displayed locally on the SLM to map out spatially varying phase differences \cite{Zupancic2016, Supikov2023}.
Here, we take a different approach and use several speckle images created by random SLM phase patterns to recover the phase, $\varphi$, and laser intensity profile, $I$, at the SLM (Fig.~\ref{fig:fast_calib_schemMain}).
Instead of modelling them using smooth analytical functions \cite{Peng2020}, we optimise each pixel value of the phase and the intensity profile using adaptive moment estimation \cite{kingma2014adam}, a variant of stochastic gradient descent (SGD).
Initially, we display a sequence of semi-random phase patterns, $\theta$, on the SLM and record the corresponding camera images, $I_{\text{cam}}$ (Fig.~\ref{fig:fast_calib_schemMain}).
The details of the experimental setup can be found in \hyperref[sec:methods]{Methods}. 
We then simulate the expected camera images, $I_{\text{sim}}$, by propagating the electric field, $E$, from the SLM plane to the image plane using a type-1 non-uniform fast Fourier transform (NUFFT) \cite{beatty2005rapid},
\begin{equation}
\label{eq:PRP}
I_{\text{sim}}=\left| \mathcal{F}\!\left\{E\right\} \right|^{2},
\end{equation}
with $E_{kl} = A_{kl}\,e^{i\left(\varphi_{kl} \, + \, \theta_{kl} \right)}$, where $A_{kl}$ is the amplitude profile of the laser beam with its intensity, $I_{kl}={\left|A_{kl}\right|}^2$, its spatially varying phase, $\varphi_{kl}$, and the phase displayed on the SLM, $\theta_{kl}$, with indices, $k$, $l$.
Using a NUFFT instead of a fast Fourier transform (FFT) allows us to choose the pixel pitch and the region of interest in the Fourier plane arbitrarily, lowering the memory requirements at the cost of execution speed.
We chose the pixel pitch in the Fourier plane to match the pixel pitch of our camera and calculated only the area of the computational image plane that is covered by the camera.
In practice, this mapping of the camera coordinates to the coordinates in the Fourier plane is not sufficiently accurate due to experimental imperfections.
To account for the mismatch between the camera image and the simulated image, we employ a partial affine transformation, $\mathcal{T}_U$, with the transformation matrix, $U$, to account for rotation, translation and scaling of the camera image \cite{Schroff2023}.
We applied this coordinate transform to compare the simulated images generated by our digital twin, $I_{\text{sim}}$, with the experimental images, $I_{\text{cam}}$, taken by the camera (Fig.~\ref{fig:fast_calib_schemMain}).
By adjusting the value of every pixel in the phase, $\varphi$, and intensity, $I$, as well as the partial affine transformation matrix, $U$, we minimise a cost function using adaptive moment estimation \cite{NEURIPS2019_9015, kingma2014adam}.
Our cost function calculates the mean squared error (MSE) to minimise the difference between the camera images and the simulated images,
\begin{equation}
C_{\text{MSE}}\!\left(\varphi,\, I,\, U\right) = \frac{1}{N_{\text{F}}N^2} \sum_{n=1}^{N_{\text{F}}} \sum_{k,\,l}^N \left[\Big(I_{\text{sim},\,kl}^{n} - \mathcal{T}_U\!\left\{{I}_{\text{cam}}^{n}\right\}_{kl} \!\Big)^2\right]\!,
\label{eq:cost}
\end{equation}
where $N_{\text{F}}$ is the number of random phase patterns used in the optimisation and $N$ the number of pixel rows and columns on the SLM ($N=1024$ for our specific SLM).
We introduce two regularisation terms, $C_{\!\varphi}$ and $C_{\!A}$, to reduce fast spatial fluctuations of the phase and the intensity profile, respectively (\hyperref[sec:methods]{Methods}).
We calculate the cost function as
\begin{equation}
C = s \left(C_{\text{MSE}} + s_{\!\varphi}C_{\!\varphi} + s_{\!A}C_{\!A} \right),
\end{equation}
with the overall steepness, $s$, and weighing parameters, $s_{\!\varphi}$ and $s_{\!A}$.
The regularisation terms ensure that the phase, $\varphi$, and the amplitude profile at the SLM, $A$, remain smooth as expected from the wavefront of a Gaussian beam and its intensity profile.
They reduce overfitting and promote global convergence by reducing the likelihood of getting stuck in a local minimum.
The cost function steepness, $s=10^{14}$, and weighing parameters $s_{\!\varphi}=5\times 10^{-3}$ and $s_{\!A}=2\times 10^{-2}$ were chosen empirically.

\subsection{Calibration of the intensity and phase}
\label{sec:constant_field}

To determine the intensity profile and phase, we run our optimisation algorithm without modelling pixel crosstalk (Fig.~\ref{fig:fast_calib_schemMain}) using smooth, semi-random phase patterns that suppress the effect of pixel crosstalk \cite{Wu2019} (details in \hyperref[sec:methods]{Methods}).
The phase, $\varphi$, is initialised with zeros and the intensity profile, $I$, with an array of ones.
After 2000 iterations using $N_{\text{F}}=10$ different random SLM phase patterns ($\sim\!\SI{7}{minutes}$ on an NVIDIA RTX A5000 GPU), we stop the optimisation as stagnation is reached.

\begin{figure}[!b] \centering{\phantomsubcaption\label{subfig:int_lpp}\phantomsubcaption\label{subfig:int_stoch}\phantomsubcaption\label{subfig:phi_lpp}\phantomsubcaption\label{subfig:phi_stoch}\phantomsubcaption\label{subfig:phi_error}}
    \includegraphics[width = 1.0\linewidth]{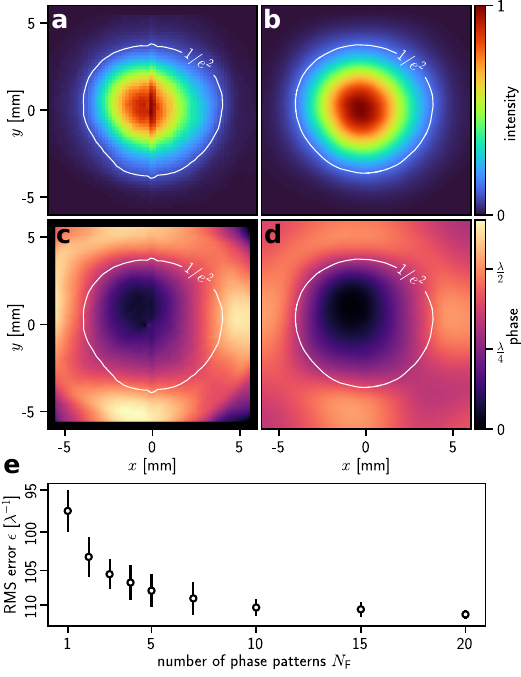}
     \caption{
     Comparison of the results generated by the local sampling method and by our stochastic approach.
     Laser intensity profiles, $I$, measured using
     (\subref{subfig:int_lpp}) the local sampling method with $64 \times 64$ local phase patterns \cite{Zupancic2016, Schroff2023}, and
     (\subref{subfig:int_stoch}) our stochastic approach with $N_{\text{F}}=20$ phase patterns.
     The region $\mathcal{B}$ lies within the $1/e^2$ intensity threshold (white line).
     (\subref{subfig:phi_lpp}) Phase at the SLM measured using the local sampling method, $\varphi_{\text{LS}}$, with $124 \times 124$ local phase patterns and
     (\subref{subfig:phi_stoch}) using our stochastic approach, $\varphi_{\text{SGD}}$, with $N_{\text{F}}=20$.
     Due to the width of the small phase patterns used in the local sampling method, the phase at the very edge of the SLM cannot be measured as indicated by the black border in (\subref{subfig:phi_lpp}).
     (\subref{subfig:phi_error}) RMS error, $\epsilon$, of the difference between the phases generated using each method, $\Delta\varphi$, as a function of the number of phase patterns, $N_{\text{F}}$, used in the stochastic approach.
     For each value of $N_{\text{F}}$, the measurement of $\varphi_{\text{SGD}}$ was repeated four times using different semi-random phase patterns each time.
     }
\label{fig:fast_calib_results}
\end{figure}

We compare the recovered phase and intensity profile to the results obtained from a calibration method that displays blazed gratings on small square regions at different positions on the SLM, locally probing $I$ and $\varphi$ at each position \cite{Zupancic2016, Schroff2023}.
We further improved the accuracy of this local sampling technique by compensating for pointing fluctuations of the incident laser beam during the measurement using a reference pattern, resulting in a residual error of $<\lambda/100$.
Comparing the intensity profile measured by the local sampling method (Fig.~\ref{subfig:int_lpp}) with the result of our stochastic approach (Fig.~\ref{subfig:int_stoch}), we find that the beam diameter calculated from the stochastic approach is $\sim\SI{9}{\%}$ larger.
The intensity profile obtained from the local sampling method contains a visible vertical line at the centre of the SLM (Fig.~\ref{subfig:int_lpp}), caused by the addressing scheme used to update the phase pattern on the LCOS SLM \cite{Hamamatsu2022}. 
This artefact is no longer present in the intensity profile of the stochastic approach since we use global SLM phase patterns instead of small, local ones.
The phases recovered using the stochastic approach (Fig.~\ref{subfig:phi_lpp}) and from the local sampling method (Fig.~\ref{subfig:phi_stoch}) agree well in the centre of the SLM, however, they deviate significantly from each other in regions on the SLM where the intensity of the laser beam is low.
To characterise $\varphi$, we only consider a region, $\mathcal{B}$, on the SLM in which the intensity is larger than $1/e^2$ of the maximum intensity.
To determine the recalibration error of the phase measurement using the stochastic method, we subtract the initially measured phase from the semi-random phase patterns and display the resulting phase, $\theta - \varphi$, on the SLM.
When re-running the stochastic method using those wavefront-corrected phase patterns, we obtain a residual phase, $\delta\varphi$, after removing the tilt within $\mathcal{B}$.
To quantify the recalibration error, we calculate the standard deviation of the residual phase
\begin{equation}
    \sigma = \sqrt{\frac{1}{N\!_{\mathcal{B}} k^2} \sum_{i,\,j \in \mathcal{B}}\left(\delta\varphi_{ij} - \overline{\delta\varphi}\right)^2},
\end{equation}
where $\overline{\delta\varphi}$ is the mean value of $\delta\varphi$ in $\mathcal{B}$, containing $N\!_\mathcal{B}$ pixels with indices $i$, $j$ and $k=2\pi/\lambda$.
With the stochastic approach ($N_{\text{F}}=10$), we obtain a similarly small standard deviation, $\sigma=\lambda/170$, compared to the local sampling method, $\sigma=\lambda / 180$.

To investigate the accuracy of the phase measured using the stochastic approach, $\varphi_{\text{SGD}}$, we quantify its deviation from the phase measured using the local sampling method, $\varphi_{\text{LS}}$, by calculating the RMS error of their difference, $\Delta\varphi = \varphi_{\text{SGD}}  - \varphi_{\text{LS}}$, in region $\mathcal{B}$,
\begin{equation}
    \epsilon = \sqrt{\frac{1}{N\!_{\mathcal{B}}k^2}\sum_{i,\, j \in\mathcal{B}} \left(\Delta\varphi_{ij} - \overline{\Delta\varphi}\right)^2},
\end{equation}
where $\overline{\Delta\varphi}$ is the mean value of $\Delta\varphi$ in $\mathcal{B}$.
Using our stochastic approach with only one SLM phase pattern ($N_{\text{F}}=1$) in the optimisation, we already obtain $\epsilon= \lambda/97$.
To investigate if the RMS error, $\epsilon$, decreases further when using more than one SLM phase pattern, we measure $\varphi_{\text{SGD}}$ using different values of $N_{\text{F}}$ and calculate $\epsilon$ for each measurement (Fig.~\ref{subfig:phi_error}).
When increasing $N_{\text{F}}$, the error decreases and reaches $\epsilon = \lambda / 110$ at $N_{\text{F}}=10$.
When the number of SLM phase patterns is increased further to $N_{\text{F}}=20$, the RMS error only decreases slightly.
Artefacts caused by the local sampling method are a significant source of error at this scale (Fig.~\ref{fig:phase_difference} in \hyperref[sec:methods]{Methods}).
The runtime of the stochastic method is much shorter and requires significantly fewer phase patterns and camera images compared to the local sampling method ($\sim\!\SI{7}{minutes}$ runtime with $N_{\text{F}} = 10$ patterns, compared to $\sim\!\SI{2.5}{hours}$ and $\sim\!20000$ patterns).
Another advantage of the stochastic approach is the spatial resolution of $1\times \SI{1}{p_{\text{SLM}}}$, with the SLM pixel pitch, $\SI{}{p_{\text{SLM}}}=\SI{12.5}{\micro\meter}$, whereas the spatial resolution of the local sampling method depends on the number of images taken ($8\times \SI{8}{p_{\text{SLM}}}$ for the phase and $16\times \SI{16}{p_{\text{SLM}}}$ for the intensity profile in this study).
For this reason, the local sampling method relies on upscaling the image of the measured phase and intensity to the native resolution of the SLM which is not required with our stochastic approach.

This increased speed and simplified method allow for the use of holographic light potentials in demanding settings such as those involving high-NA objectives or other optical elements that distort the beam profile \cite{gross2021quantum, SchroffConference2023}.
At higher laser beam intensities, the SLM surface can deform due to thermal lensing, causing the calibration to depend on power \cite{matsumoto2014stable}.
With a calibration time of only a few minutes, recalibration is possible in such scenarios.
This makes it easier to optimise light potentials using lightshifts on ultracold atoms in quantum-gas microscopes and neutral-atom tweezer arrays, as fewer iterations and images are needed overall.

\subsection{Modelling pixel crosstalk}
\label{sec:PCT}

The effect of pixel crosstalk is most noticeable for light potentials with high spatial frequencies in the SLM phase pattern, containing many $0$ to $2 \pi$ phase jumps which are affected most by pixel crosstalk.
To model pixel crosstalk, one generally convolves the displayed phase pattern with a kernel that mimics the non-discrete phase changes between neighbouring SLM pixels \cite{Moreno2021, Pushkina2020, Schroff2023, Ronzitti2012, Buske2024, Jang2024}.
Recently, a fast crosstalk model that avoids the computationally expensive convolution has been proposed \cite{Moser2019}.
Finding the optimal parameters for the pixel crosstalk model has previously been achieved by an exhaustive search approach, where a small number of parameters were optimised to obtain the best agreement between a simulated image and the corresponding camera image \cite{Schroff2023, Buske2024}.
More sophisticated pixel crosstalk models have been developed \cite{Moser2019}, however, optimising their large parameter space via a direct search is infeasible.
Alternatively, pixel crosstalk parameters have been found by measuring higher diffraction orders \cite{Ronzitti2012, Moser2019, Moreno2021} which are not accessible in our experiment as they lie outside the field of view of our camera.

\begin{figure}[!]
    \centering{\phantomsubcaption\label{subfig:kernel1}\phantomsubcaption\label{subfig:kernel2}\phantomsubcaption\label{subfig:kernel3}\phantomsubcaption\label{subfig:kernel4_P3}\phantomsubcaption\label{subfig:kernel4_P5}\phantomsubcaption\label{subfig:kernel4_P7}}
    \includegraphics[width = 1.0\linewidth]{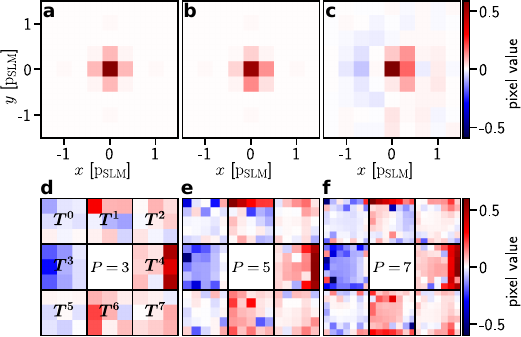}
    \caption{
    Optimised pixel crosstalk models. (\subref{subfig:kernel1}-\subref{subfig:kernel3}) Convolution kernels found using the optimisation, normalised by their pixel sum. 
    (\subref{subfig:kernel1}) Kernel, $K_{\mathrm{I}}$, found using Model~$\boldsymbol{\mathrm{I}}$ (Eq.~\ref{eq:ct_kernel1}) with parameters $q=1.20$ and $\sigma = \SI{2.03}{p_{\text{SLM}}}$.
    (\subref{subfig:kernel2}) Kernel, $K_{\mathrm{II}}$, after optimising Model~$\boldsymbol{\mathrm{II}}$ (Eq.~\ref{eq:kernel_pw}) resulting in $q_{\text{xp}}=1.85$, $q_{\text{xn}}=1.15$, $q_{\text{yp}}=1.75$, $q_{\text{yn}}=1.12$ and  $\sigma_{\text{xp}}=\SI{1.02}{p_{\text{SLM}}}$, $\sigma_{\text{xn}}=\SI{2.72}{p_{\text{SLM}}}$, $\sigma_{\text{yp}}=\SI{1.06}{p_{\text{SLM}}}$, $\sigma_{\text{yn}}=\SI{3.04}{p_{\text{SLM}}}$.
    (\subref{subfig:kernel3}) Optimised kernel, $K_{\mathrm{III}}$, using Model~$\boldsymbol{\mathrm{III}}$, where each pixel is a learnable parameter. 
    (\subref{subfig:kernel4_P3}-\subref{subfig:kernel4_P7}) Optimised $T^i$ of Model~$\boldsymbol{\mathrm{IV}}$ (Eq.~\ref{eq:moser_model}) with upscaling factors (\subref{subfig:kernel4_P3}) $P=3$, (\subref{subfig:kernel4_P5}) $P=5$, and (\subref{subfig:kernel4_P7}) $P=7$.     
            }
\label{fig:PXT_optimised}
\end{figure}
Here, we investigate four different pixel crosstalk models (details in \hyperref[sec:methods]{Methods}).
Models~$\boldsymbol{\mathrm{I}}$, $\boldsymbol{\mathrm{II}}$ and $\boldsymbol{\mathrm{III}}$ convolve the displayed phase pattern with a kernel parameterised in three different ways: 
Model~$\boldsymbol{\mathrm{I}}$ considers a radially symmetric super-Gaussian kernel, $K_{\mathrm{I}}$ (Eq.~\ref{eq:ct_kernel1}).
Model~$\boldsymbol{\mathrm{II}}$ uses a piecewise super-Gaussian kernel, $K_{\mathrm{II}}$, to allow for asymmetries in the $x$ and $y$ directions (Eq.~\ref{eq:kernel_pw}).
Finally, Model~$\boldsymbol{\mathrm{III}}$ treats every pixel in the kernel, $K_{\mathrm{III}}$, as a learnable parameter.
The SLM phase after modelling pixel crosstalk is then calculated using a convolution (Eq.~\ref{eq:conv}) after upscaling the SLM phase pattern by a factor $P$ using nearest-neighbour interpolation.
In addition to the three convolutional models, we investigate the fast Model~$\boldsymbol{\mathrm{IV}}$, based on previous work \cite{Moser2019}.
Model~$\boldsymbol{\mathrm{IV}}$ utilises eight matrices, $T^i$ with $i\in0, 1, ..., 7$, that each correspond to the crosstalk caused by the eight neighbouring pixels (Eq.~\ref{eq:moser_model} and Fig.~\ref{fig:moser_model}).
The crosstalk of each SLM pixel is calculated in parallel on the GPU, making this model faster to compute than a convolution.
In contrast to previous work \cite{Moser2019}, where each $T^i$ was constructed from one-dimensional analytical functions, we treat each of the $P \times P$ elements in every $T^i$ as a learnable parameter.

\begin{figure*}[ht!] 	
\centering{\phantomsubcaption\label{subfig:square_potential}\phantomsubcaption\label{subfig:convergence}\phantomsubcaption\label{subfig:square_profile}\phantomsubcaption\label{subfig:square_size}}
    \includegraphics[width=0.9\linewidth]{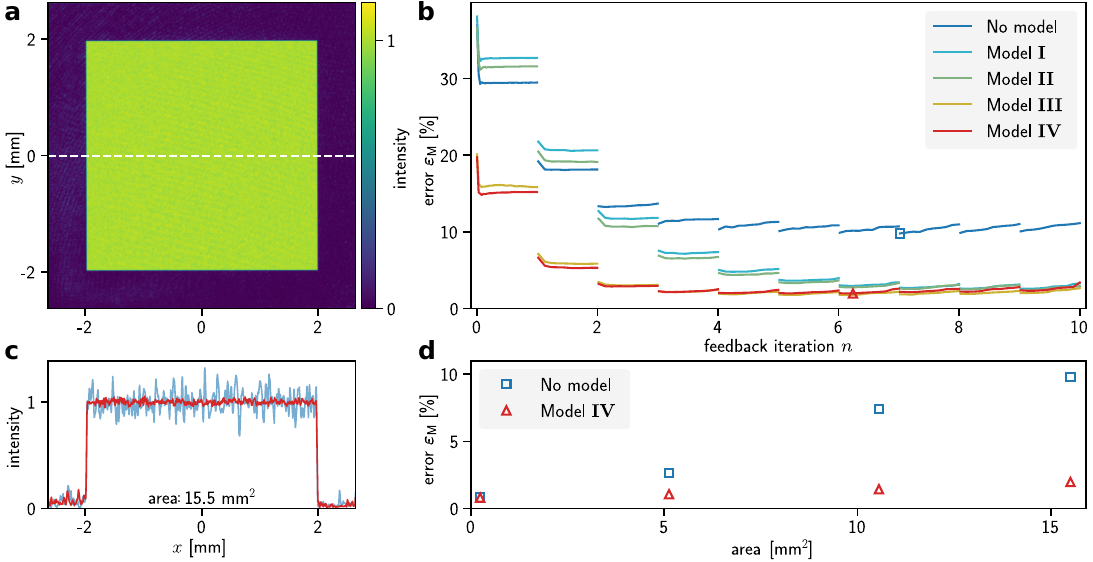}
    \caption{
    Benchmark of different crosstalk models.
    (\subref{subfig:square_potential}) Large $\SI{15.5}{mm^2}$ top-hat potential generated using Model~$\boldsymbol{\mathrm{IV}}$ after $n=6$ feedback iterations.
    (\subref{subfig:convergence}) Convergence of the RMS error, $\varepsilon_{\text{M}}$, during the camera feedback process as a function of feedback iterations, $n$, using pixel crosstalk Models $\boldsymbol{\mathrm{I}}$~-~$\boldsymbol{\mathrm{IV}}$.
    The lowest error reached without modelling pixel crosstalk (blue square) is $\sim5$ times larger than the lowest error reached using pixel crosstalk Model~$\boldsymbol{\mathrm{IV}}$ (red triangle).
    (\subref{subfig:square_profile}) Profiles of the top-hat potential in (\subref{subfig:square_potential}) along the white dashed line using crosstalk Model~$\boldsymbol{\mathrm{IV}}$ (red line) and without modelling crosstalk (blue line).
    (\subref{subfig:square_size}) The RMS error, $\varepsilon_{\text{M}}$, after camera feedback increases as a function of the area of the top-hat potential.
    }
\label{fig:rmse_crosstalk}	
\end{figure*}

To optimise each of the four models, we include them in our digital twin of the experiment (Fig.~\ref{fig:fast_calib_schemMain}).
We generate entirely random SLM phase patterns with uniformly distributed values between $0$ and $2\pi$ on each SLM pixel, increasing the effect of pixel crosstalk compared to the smooth phase patterns used in Section~\ref{sec:constant_field}.
After applying the pixel crosstalk model, we upscale the laser amplitude profile, $A$, and the phase, $\varphi$, by a factor $P$ using Lanczos interpolation \cite{Lanczos1938} to match the resolution of the upscaled SLM phase, $\Theta$.
Then, the camera images, $I_{\text{cam}}$, are simulated by propagating the upscaled electric field from the SLM to the image plane via a NUFFT (Fig.~\ref{fig:fast_calib_schemMain}).
Similar to Section~\ref{sec:constant_field}, we optimise the parameters of the pixel crosstalk model using adaptive moment estimation by minimising the mean-squared error between camera images, $I_{\text{cam}}$, and the corresponding simulated images, $I_{\text{sim}}$, using the cost, $C_{\text{CT}}=s\,C_{\text{MSE}}$.
For $N_{\text{F}}=10$ phase patterns with an upscaling factor $P=3$, the optimisation converges after $\sim300$ iterations with a runtime of $\sim4$ minutes for Models $\boldsymbol{\mathrm{I}}$-$\boldsymbol{\mathrm{III}}$ and $\sim2$ minutes for Model~$\boldsymbol{\mathrm{IV}}$ using our GPU.

Optimising Model~$\boldsymbol{\mathrm{I}}$ results in similar parameters compared to previous work using the same LCOS SLM \cite{Pushkina2020, Schroff2023} (Fig.~\ref{subfig:kernel1}).
We obtain a similar kernel using Model~$\boldsymbol{\mathrm{II}}$, with a slight asymmetry along the $x$ and $y$ directions (Fig.~\ref{subfig:kernel2}).
Both models were initialised with parameters $q_i=2$ and $\sigma_i=\SI{1}{p_{\text{SLM}}}$ (\hyperref[sec:methods]{Methods}).
In the optimised kernel of Model~$\boldsymbol{\mathrm{III}}$, horizontal asymmetry is visible with negative pixel values in the left half (Fig.~\ref{subfig:kernel3}).
We initialised this kernel by setting the central pixel of $K_{\mathrm{III}}$ to one and the remaining pixels to zero.
For Model $\boldsymbol{\mathrm{IV}}$, we investigate different upscaling factors, $P=3$, $P=5$, and $P=7$.
Each $T^i$ is initialised by setting all elements to zero.
The optimised $T^i$ show noticeable asymmetry along the horizontal axis that is consistent for different values of $P$ (Fig.~\ref{subfig:kernel4_P3}-\subref{subfig:kernel4_P7}).
For each value of $P$, we repeat the optimisation three times using different phase patterns and camera images during each run, resulting in 
very similar $T^i$ for $P=3$, with their standard deviation not exceeding $\SI{1}{\%}$ (Fig.~\ref{fig:kernel_moser_SI} in \hyperref[sec:methods]{Methods}).
When increasing $P$, the standard deviation of individual elements only rises marginally due to the larger number of parameters.
We conclude that the information provided by $N_{\text{F}}=10$ phase patterns and camera images is sufficient to train crosstalk Model~$\boldsymbol{\mathrm{IV}}$, containing $392$ free parameters at $P=7$.

\subsection{Light potential quality and efficiency}
\label{sec:crosstalk_benchmark}

To benchmark the performance of the different pixel crosstalk models and to show that our calibration method produces accurate light potentials, we generate square, top-hat-shaped light potentials (Fig.~\ref{subfig:square_potential}) using conjugate gradient (CG) minimisation \cite{Harte2014} and a camera feedback algorithm \cite{Schroff2023, Bruce2015} with 10 camera feedback iterations and 50 CG iterations each.
We investigate the convergence of the RMS error \cite{Schroff2023}, $\varepsilon_{\text{M}}$, of the light potential generated by each crosstalk model (Fig.~\ref{subfig:convergence}) and measure the efficiency, $\eta_{\text{M}}$ (Table~\ref{tab:ModelsTable}, details in \hyperref[sec:methods]{Methods}).
Without modelling pixel crosstalk, the RMS error of the square, top-hat shaped light potential reaches its minimum of $\SI{9.8}{\%}$ after $n=7$ feedback iterations (blue square in Fig.~\ref{subfig:convergence}).
Using Model $\boldsymbol{\mathrm{I}}$ and $\boldsymbol{\mathrm{II}}$, this error reduces to $\SI{2.6}{\%}$ and $\SI{2.4}{\%}$, respectively. 
Interestingly, the error before any camera feedback ($n=1$) using Models~$\boldsymbol{\mathrm{I}}$ and $\boldsymbol{\mathrm{II}}$ is slightly larger than without modelling crosstalk.
Model~$\boldsymbol{\mathrm{III}}$ results in a significantly lower RMS error before any camera feedback ($\varepsilon_{\text{M}}=\SI{16}{\%}$ compared with $\SI{29}{\%}$).
Furthermore, the error using Model~$\boldsymbol{\mathrm{III}}$ converges at a faster rate compared with Models~$\boldsymbol{\mathrm{I}}$ and $\boldsymbol{\mathrm{II}}$.
An error of $\SI{2.4}{\%}$ is reached after $n=4$ feedback iterations which required $n=9$ feedback iterations using Model~$\boldsymbol{\mathrm{II}}$.
The lowest error reached with Model~$\boldsymbol{\mathrm{III}}$ is $\varepsilon_{\text{M}}=\SI{1.8}{\%}$ after $n=6$ feedback iterations.
We find that for this model, increasing the upscaling factor from $P=3$ to $P=5$ does not further reduce the RMS error.
Using Model~$\boldsymbol{\mathrm{IV}}$, the error converges at a rate similar to that of Model~$\boldsymbol{\mathrm{III}}$.
Before any camera feedback, the RMS error is slightly lower and the lowest error, $\varepsilon_{\text{M}}=\SI{2.0}{\%}$, is reached after $n=6$ iterations.
Here, increasing the upscaling factor to $P=5$ shows a small improvement (Table~\ref{tab:ModelsTable}), however, using $P=7$ did not further reduce the RMS error.
\begin{table}[t]
\centering
\renewcommand{\arraystretch}{1}
    \begin{tabular}{@{}lcccc@{}}
        \toprule
        Crosstalk model & $P$ & \multicolumn{2}{c}{$\varepsilon_{\text{M}}$ [\%]} & $\eta_{\text{M}}$ [\%] \\ \cmidrule{3-4}
         &       & $n=1$                       &  $n=n_{\text{min}}$                       & \\
        \midrule
        Without                             & 1 & 29.5 & 9.8 & 14    \\
        \addlinespace
        Model $\boldsymbol{\mathrm{I}}$     & 3 & 32.7 & 2.6 & 11    \\
        Model $\boldsymbol{\mathrm{II}}$    & 3 & 32.6 & 2.4 & 11    \\
        Model $\boldsymbol{\mathrm{III}}$   & 3 & 15.1 & 1.8 & 13    \\
        Model $\boldsymbol{\mathrm{IV}}$    & 3 & 15.8 & 2.0 & 14    \\
        \addlinespace
        Model $\boldsymbol{\mathrm{III}}$   & 5 & 15.8 & 2.0 & 14    \\
        Model $\boldsymbol{\mathrm{IV}}$    & 5 & 13.9 & 1.9 & 14    \\
        \bottomrule
    \end{tabular}
    \caption{Summary of results presented in Fig.~\ref{fig:rmse_crosstalk}.
    A top-hat potential with a was generated using crosstalk Models $\boldsymbol{\mathrm{I}}$~-~$\boldsymbol{\mathrm{IV}}$ with an upscaling factor of $P=3$.
    Models $\boldsymbol{\mathrm{III}}$~and~$\boldsymbol{\mathrm{IV}}$ were also investigated using $P=5$.
    The lowest RMS error, $\varepsilon_{\text{M}}$, reached during the camera feedback process (at $n=n_{\text{min}}$) and before any camera feedback ($n=1$) is shown.
    }
    \label{tab:ModelsTable}
\end{table}

In Fourier imaging setups, the effect of pixel crosstalk becomes particularly noticeable for large light potentials which require high spatial frequencies to be displayed on the SLM \cite{Schroff2023}.
To investigate this effect, we vary the width of the square target potential from $\sim\!\SI{0.5}{mm}$ to $\sim\!\SI{4.0}{mm}$
and perform 10 camera feedback iterations for each square without modelling pixel crosstalk.
To obtain similar efficiencies for the differently sized potentials, we vary the curvature, $R$, of the quadratic initial phase guess \cite{Schroff2023} linearly with the area of the potential, from $R=\SI{0.2}{mrad\per {p_{\text{SLM}}} \squared}$ to $R=\SI{1.6}{mrad\per {p_{\text{SLM}}} \squared}$ (Fig.~\ref{subfig:square_size}).
We then repeat this measurement with crosstalk Model~$\boldsymbol{\mathrm{IV}}$ using an upscaling factor of $P=3$.
The RMS error increases linearly as a function of the light potential's area for both sets of measurements, however, at different rates.
For the smallest square potential, modelling the pixel crosstalk does not reduce the RMS error below $\varepsilon_{\text{M}}\sim\SI{0.8}{\%}$.
The RMS error of the largest square potential is reduced from $\sim \SI{10}{\%}$ without modelling pixel crosstalk to $\sim\SI{2}{\%}$ using Model~$\boldsymbol{\mathrm{IV}}$.

Without modelling pixel crosstalk, the experimentally measured efficiency of the largest top hat is $\eta_{\text{M}}\sim\SI{14}{\%}$ (\hyperref[sec:methods]{Methods}).
When Models~$\boldsymbol{\mathrm{I}}$ and $\boldsymbol{\mathrm{II}}$ are used, the efficiency of the potential drops to $\eta_{\text{M}}\sim\SI{11}{\%}$, which is consistent with our previous findings \cite{Schroff2023}.
Models~$\boldsymbol{\mathrm{III}}$ and $\boldsymbol{\mathrm{IV}}$ do not significantly decrease the efficiency of the potentials, resulting in efficiencies of $\eta_{\text{M}}\sim\SI{13}{\%}$ and $\eta_{\text{M}}\sim\SI{14}{\%}$, respectively.
The efficiencies of the top hats are relatively low since a large curvature, $R$, of the initial phase guess was necessary to prevent the formation of optical vortices in the light potential during the feedback process.
This large phase curvature causes a significant loss of optical power to areas outside the signal region \cite{Schroff2023}.
Employing a vortex-removal technique \cite{Schroff2023} or adding a phase gradient term to the cost function when solving the phase retrieval problem \cite{Choi2021} would allow the use of smaller values of $R$ in the initial phase guess, increasing the efficiency of the light potential.
Alternatively, an efficiency term can be added to the cost function to optimise the optical power in the potential \cite{Ebadi2021, chao_2023}.

\begin{figure}[!b]
\centering{\phantomsubcaption\label{subfig:spot_array}\phantomsubcaption\label{subfig:top_hat}\phantomsubcaption\label{subfig:zoom}}
\includegraphics[width = 1.0\linewidth]{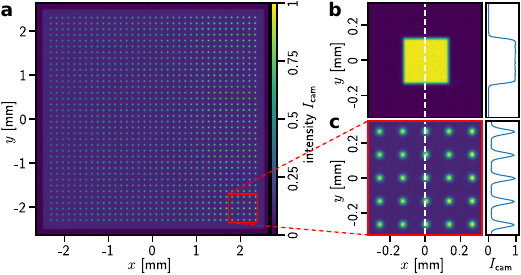}
\caption{Light potentials generated using camera feedback with crosstalk Model~$\boldsymbol{\mathrm{IV}}$.
(\subref{subfig:spot_array})~Spot array containing $36\times36$ Gaussian spots with an RMS error, $\varepsilon_{\text{M}}=\SI{1.6}{\%}$, and efficiency, $\eta_{\text{M}}=\SI{11}{\%}$.
(\subref{subfig:top_hat}) Small top-hat potential with $\varepsilon_{\text{M}}=\SI{0.9}{\%}$ and $\eta_{\text{M}}=\SI{39}{\%}$.
(\subref{subfig:zoom}) Close-up of the potential within the red box in (\subref{subfig:spot_array}).
The insets in (\subref{subfig:top_hat}-\subref{subfig:zoom}) show the profiles of the potentials along the dashed white lines.}
\label{fig:ShowPotentials}
\end{figure}

To demonstrate that our protocol can produce more complex light potentials, we generated a square array with $1296$ Gaussian spots on a small, constant background measuring \SI{5}{mm} across.
Additionally, the spot intensity is varied in each column, increasing from left to right.
After $n=6$ iterations using Model~$\boldsymbol{\mathrm{IV}}$ with $P=3$, the RMS error of the spot array reached $\varepsilon_{\text{M}}=\SI{1.6}{\%}$ with an efficiency of $\eta_{\text{M}}=\SI{11}{\%}$.
A few optical vortices are present in the spot array which could be eliminated using a vortex removal technique \cite{Schroff2023}.
The standard deviation of the spot intensities in the most intense column of spots is especially low at \SI{0.7}{\%} which is sufficient for tunnel-coupled tweezers in Hubbard arrays \cite{young2022tweezer,spar2022realization}.
Using our protocol, we also generated a smaller ($\SI{0.06}{mm^2}$) top-hat potential (Fig.~\ref{subfig:top_hat}) with an especially high efficiency $\eta_{\text{M}}=\SI{39}{\%}$ and low error, $\varepsilon_{\text{M}}=\SI{0.9}{\%}$.

\section{Conclusion}
\label{sec:conclusion}
In conclusion, we used adaptive moment estimation to measure the laser intensity profile and the phase at the SLM in a few minutes using 1-10 camera images while maintaining a low recalibration error of $\sim\lambda/170$ which is on par with methods that use thousands of images.
In a second step, the crosstalk between neighbouring pixels on the SLM was characterised using a computationally efficient model (Model~$\boldsymbol{\mathrm{IV}}$) and compared to models that use a convolution (Models~$\boldsymbol{\mathrm{I}}$-$\boldsymbol{\mathrm{III}}$).
We then benchmarked the accuracy of these models by using them to generate a large, top-hat-shaped potential.
The computationally efficient Model~$\boldsymbol{\mathrm{IV}}$ and the model using a convolution kernel without any analytical constraints (Model~$\boldsymbol{\mathrm{III}}$) converged rapidly to an error of $\varepsilon_{\text{M}}\sim\SI{2}{\%}$. 
Due to the low number of camera images needed, this calibration technique paves the way towards in situ optimisation of light potentials in cold atom experiments using optical lattices or tweezer arrays.
Extending our approach to high-NA optics using more rigorous light propagation, such as the angular spectrum method \cite{goodman2017}, remains the subject of future work.
Our method for calibrating a computationally efficient pixel crosstalk model will benefit applications in holographic displays by reducing inhomogeneities in the generated images, particularly for displays with a large field of view that require large diffraction angles to be displayed on the SLM.

\section{Methods}
\label{sec:methods}

\subsection{Experimental setup}  \label{sec:methods_Setup}
We use an SLM (Hamamatsu X13138-07, $\SI{}{p_{\text{SLM}}}=\SI{12.5}{\micro\meter}$ pixel pitch, $1272 \times 1024$ pixels) in a Fourier imaging setup, similar to our previous work \cite{Schroff2023}.
A laser beam ($\lambda=\SI{670}{\nano\meter}$ wavelength) from a single-mode fibre is collimated using a triplet lens (Melles Griot 06 GLC 001) and polarised horizontally by passing it through a polarising beam splitter.
The beam is expanded using a telescope (Thorlabs GBE10-B) to a beam diameter of $\sim\SI{7.4}{mm}$ and is incident onto the SLM at an angle of $\sim10^{\circ}$.
The light reflected by the SLM passes through a 2" doublet lens with focal length $f=\SI{250}{mm}$ (Thorlabs ACT508-250-B) to form an image at the camera (Andor Zyla 5.5, $\SI{}{p_{\text{CAM}}}=\SI{6.5}{\micro\meter}$ pixel pitch, $2560 \times 2160$ pixels) in its Fourier plane.
At $\lambda = \SI{670}{nm}$, our SLM can display phases of up to $4\pi$.

\subsection{Implementation of the stochastic optimisation} \label{sec:methods_Cali}

The semi-random phase patterns used in Section~\ref{sec:constant_field} are generated from a $128\times128$ array containing random values between $0$ and $3\pi$.
This array is upscaled to the resolution of the SLM (we use the central $N\times N$ pixels on the SLM with $N=1024$) using nearest neighbour interpolation and convolved with a Gaussian kernel of \SI{8}{p_{\text{SLM}}} width, producing smooth phase patterns 
with phase values between $0$ and $\sim\!2.5\pi$.

We calculate the two regularisation terms, $C_{\varphi}$ and $C_A$, to reduce the gradient of the phase and the intensity profile at the SLM using the forward difference,
\begin{align}
	C_{\varphi} &= \frac{1}{\left(\!N\!-\! 1\!\right)^{2}}\!\sum_{i,\,j}^{N-1}\! \left[\!\left(\varphi_{i,\,j}\!-\!\varphi_{i+1,\,j}\right)^{\!2}\!+\! \left(\varphi_{i,\,j}\!-\!\varphi_{i,\,j+1}\right)^{\!2}\right]\, \text{and}\\
    C_{A} &=       \frac{1}{\left(\!N\!-\! 1\!\right)^{2}}\!\sum_{i,\,j}^{N-1}\!\left[\!\left(A_{ij}\!-\!A_{i+1,\,j}\right)^{\!2}\!+\! \left(A_{i,\,j}\!-\!A_{i,\,j+1}\right)^{\!2}\right].
\end{align}

To determine the NUFFT coordinates of the camera pixels in the computational Fourier plane in the range $[-\pi,\, \pi)$, we find the position of the zeroth-order diffraction spot on the camera by fitting a Gaussian to it.
We then convert the pixel pitch of the camera in the Fourier plane to radians, $\SI{}{p_{\text{NUFFT}}} = \frac{2\pi}{\lambda f}\SI{}{p_{\text{SLM}}} \SI{}{p_{\text{CAM}}}$.

\subsection{Characterising the phase measurement}

\begin{figure}[b]
\centering{\phantomsubcaption\label{subfig:rep_lpp}\phantomsubcaption\label{subfig:rep_stoch}}
    \includegraphics[width = 1.0\linewidth]{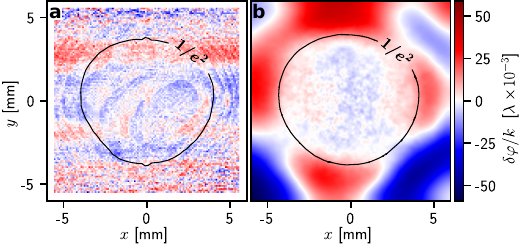}
     \caption{
      Residual phase, $\delta\varphi$, obtained from (\subref{subfig:rep_lpp}) the local sampling method and (\subref{subfig:rep_stoch}) the stochastic approach with $N_{\text{F}}=20$.   }
\label{fig:phase_repeatability}
\end{figure}

\begin{figure}[t]
\centering{\phantomsubcaption\label{subfig:nf1}\phantomsubcaption\label{subfig:nf3}\phantomsubcaption\label{subfig:nf20}}
    \includegraphics[width = 1.0\linewidth]{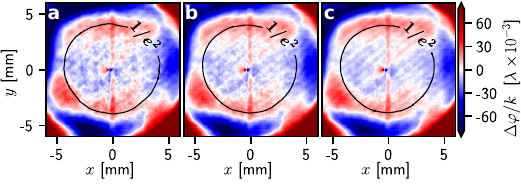}
     \caption{Phase difference, $\Delta\varphi$, between the phase measured obtained from the local sampling method and from the stochastic approach using (\subref{subfig:nf1}) $N_\text{F}=1$, (\subref{subfig:nf3}) $N_\text{F}=3$, and (\subref{subfig:nf20}) $N_\text{F}=20$ SLM phase patterns.
     The RMS error, $\epsilon$, is calculated within the $1/e^2$ region, $\mathcal{B}$ (black line).}
\label{fig:phase_difference}
\end{figure}

To characterise the recalibration error, $\sigma$, of the local sampling phase measurement and the stochastic approach we study the residual phase, $\delta\varphi$. 
The residual phase is obtained after running each method twice, where the result from the first run is used to compensate for the wavefront in the second run, producing a flat phase with a residual error (Fig.~\ref{fig:phase_repeatability}). \\
To determine the agreement between the phases measured using the local sampling method, $\varphi_{\text{LS}}$, and using the stochastic approach, $\varphi_{\text{SGD}}$, we calculate the RMS error, $\epsilon$, from their difference, $\Delta\varphi$ (Fig.~\ref{fig:phase_difference} and Fig.~\ref{subfig:phi_error}).

\subsection{Pixel crosstalk models}

\begin{figure}[b]\centering{\phantomsubcaption\label{subfig:ct_neighbours}\phantomsubcaption\label{subfig:ct_tarrays}}
    \includegraphics[width = 1.0\linewidth]{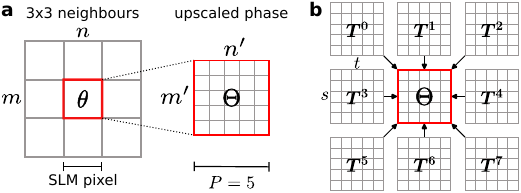}
    \caption{Pixel crosstalk Model~$\boldsymbol{\mathrm{IV}}$.
    (\subref{subfig:ct_neighbours}) The pixel crosstalk on each SLM pixel (red square) is calculated from its $3\times3$ neighbourhood, upscaling it by a factor $P$.
    (\subref{subfig:ct_tarrays}) The upscaled displayed phase, $\Theta$, of the central SLM pixel (red square) within each $3\times3$ neighbourhood is constructed by a weighted sum of eight transition matrices, $T^i$, that model the crosstalk between the central pixel and its eight neighbouring pixels.
    }
\label{fig:moser_model}
\end{figure}

The pixel crosstalk is modelled on a sub-pixel scale by upscaling the SLM phase pattern, $\theta$, using nearest-neighbour interpolation and convolving it with a parameterised crosstalk kernel \cite{Pushkina2020, Moreno2021, Ronzitti2012, Buske2024, Jang2024},
\begin{equation}  
    \Theta\!\left(x,\,y\right) = \theta\!\left(x,\,y\right) \circledast K\!\left(x,\,y\right),
\label{eq:conv}
\end{equation}
where $\Theta$ is the phase after applying the pixel crosstalk model. 

Model $\boldsymbol{\mathrm{I}}$: The crosstalk kernel is often calculated using a symmetric super-Gaussian \cite{Moser2019, Pushkina2020, Moreno2021, Schroff2023},
\begin{equation}
\label{eq:ct_kernel1}
    K_{\mathrm{I}}\!\left(x, y\right) =  \mathcal{F}^{-1} \left\{\exp \!\left[ - \!\left(\frac{\left|\kappa_{x}\right|^q + \left|\kappa_{y}\right|^q}{\sigma^q}\!\right) \!\right] \right\},
\end{equation}
with width, $\sigma$, order, $q$, and spatial frequencies, $\kappa_x$ and $\kappa_y$.

Model $\boldsymbol{\mathrm{II}}$: To capture possible asymmetries, we introduce a piecewise crosstalk kernel $K_{\mathrm{II}}$ \cite{Moser2019},

\begin{equation}
\label{eq:kernel_pw}
\begin{aligned}
\hspace{-0mm}K_{\mathrm{II}}\!\left(x, y\right)\! = \!
\begin{cases} \!\mathcal{F}^{-1}\!\left\{\!\exp\!\left[-\!\left(\!\frac{\left|\kappa_{x}\right|}{\sigma_{\text{xn}}}\!\right)^{\!q_{\text{xn}}}\!-\!\left(\!\frac{|\kappa_{y}|}{\sigma_{\text{yn}}}\!\right)^{\!q_{\text{yn}}}\right]\!\right\} &\hspace{-2mm} x\!\leq\!0,\, y\!\leq\!0\\
\!\mathcal{F}^{-1}\!\left\{\!\exp\!\left[-\!\left(\!\frac{\left|\kappa_{x}\right|}{\sigma_{\text{xn}}}\!\right)^{\!q_{\text{xn}}}\!-\!\left(\!\frac{|\kappa_{y}|}{\sigma_{\text{yp}}}\!\right)^{\!q_{\text{yp}}}\right]\!\right\} &\hspace{-2mm} x\!\leq\!0,\, y\!>\!0\\
\!\mathcal{F}^{-1}\!\left\{\!\exp\!\left[-\!\left(\!\frac{\left|\kappa_{x}\right|}{\sigma_{\text{xp}}}\!\right)^{\!q_{\text{xp}}}\!-\!\left(\!\frac{|\kappa_{y}|}{\sigma_{\text{yn}}}\!\right)^{\!q_{\text{yn}}}\right]\!\right\} &\hspace{-2mm} x\!>\!0,\, y\!\leq\!0\\
\!\mathcal{F}^{-1}\!\left\{\!\exp\!\left[-\!\left(\!\frac{|\kappa_{x}|}{\sigma_{\text{xp}}}\!\right)^{\!q_{\text{xp}}}\!-\!\left(\!\frac{|\kappa_{y}|}{\sigma_{\text{yp}}}\!\right)^{\!q_{\text{yp}}}\right]\!\right\} &\hspace{-2mm} x\!>\!0,\, y\!>\!0\\
\end{cases}
\end{aligned}
\end{equation}
with different orders, $q_{\text{i}}$, and widths, $\sigma_{\text{i}}$, along the $x$ and $y$ directions for each quadrant of the crosstalk kernel.

Model $\boldsymbol{\mathrm{III}}$: This convolutional model uses an entirely unconstrained kernel, $K_{\mathrm{III}}$, where each pixel value is a learnable parameter.
The convolution kernels $K_{\mathrm{I}}$-$K_{\mathrm{III}}$ are $3\times3$ SLM pixels wide and contain $(3P)^2$ elements each. 

\begin{figure}[t]
\centering{\phantomsubcaption\label{subfig:moser_std3}\phantomsubcaption\label{subfig:moser_std5}\phantomsubcaption\label{subfig:moser_std7}}
    \includegraphics[width = 1.0\linewidth]{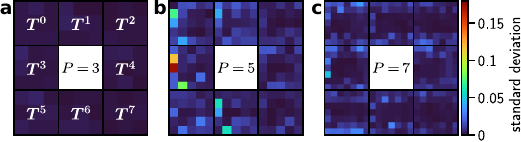}
    \caption{Standard deviation of matrices $T^i$ in Model $\boldsymbol{\mathrm{IV}}$ for (\subref{subfig:moser_std3}) $P=3$, (\subref{subfig:moser_std5}) $P=5$ and (\subref{subfig:moser_std7}) $P=7$, after performing the optimisation three times for each value of $P$ using different sets of phase patterns and camera images. The matrices $T^i$ are shown in Fig.~\ref{fig:PXT_optimised}.}
\label{fig:kernel_moser_SI}
\end{figure}

Model $\boldsymbol{\mathrm{IV}}$: This model calculates $\Theta$ directly (Fig.~\ref{fig:moser_model}) \cite{Moser2019},
\begin{equation}
\label{eq:moser_model}
\begin{aligned}
    \Theta_{m',\,n'} \!=\! \theta_{m,\,n} & \!+\! T^0_{s,\,t} (\theta_{m-1,\,n-1} \!-\! \theta_{m,\,n}\!) \!+\! T^1_{s,\,t} (\theta_{m-1,\,n} \!-\! \theta_{m,\,n}\!) \\
    & \!+\! T^2_{s,\,t} (\theta_{m-1,\,n+1} \!-\! \theta_{m,\, n}\!) \!+\! T^3_{s,\,t} (\theta_{m,\,n-1} \!-\! \theta_{m,\,n}\!) \\
    & \!+\! T^4_{s,\,t} (\theta_{m,\,n+1} \!-\! \theta_{m,\,n}\!) \!+\!T^5_{s,\,t} (\theta_{m+1,\,n-1} \!-\! \theta_{m,\,n}\!) \\
    & \!+\! T^6_{s,\,t} (\theta_{m+1,\, n} \!-\! \theta_{m,\,n}\!) \!+\! T^7_{s,\,t} (\theta_{m+1,\,n+1} - \theta_{m,\,n}\!),\\
\end{aligned}
\end{equation}
with the indices of the upscaled phase $m',\,n'\in 0,\,1,\,...,\,PN$ and the indices of the SLM pixels $m=\left\lfloor{m'/P}\right\rfloor$ and $n=\left\lfloor{n'/P}\right\rfloor$.
The matrices $T^i_{s,\,t}$ with indices $s=\text{mod}\!\left(m',\,P\right)$ and $t=\text{mod}\!\left(n',\,P\right)$, each correspond to the crosstalk caused by the eight neighbouring pixels, each containing $P \times P$ pixels, with the upscaling factor, $P$.
Each pixel value in the arrays $T^i$ is a learnable parameter. 
The difference between our model and the previous implementation \cite{Moser2019} is that each pixel $T^i$ is a free parameter.
In the original model, the two-dimensional $T^i$ were constructed from analytical one-dimensional transition functions with varying parameters depending on the pixel values of two neighbouring pixels.
Fig.~\ref{fig:kernel_moser_SI} shows the standard deviation of the measured transition matrices $T^i$ for three upscaling factors $P= 3, 5, 7$.

\subsection{Target light potential}
To characterise the quality of the potentials, we define the measured RMS error as follows:
\begin{equation}
\varepsilon_{\text{M}} = \sqrt{\frac{1}{N_{U}}\sum_{u,v\in M_{U}}\frac{\left(\hat{T}_{uv}-\hat{I}_{uv}\right)^2}{\hat{T}_{uv}^2}}\,,
\label{eq:rmse}
\end{equation}
which evaluates the camera image, $\hat{I}_{uv}$, in a measure region, $M_U$, which is defined as the region in the image plane where the target intensity, $\hat{T}_{uv}$, is greater than \SI{50}{\%} of the maximum target intensity \cite{Bruce2015, Schroff2023}, containing $N_U$ pixels with row and column indices on the camera $u$, $v$. 
We define the experimental efficiency of the light potential, $\eta_{\text{M}}$, as the ratio of the optical power, $P_S$, in the transformed signal region, $S_U$, to the measured power of the beam, $P_{\text{in}}$ \cite{Schroff2023}: 
\begin{equation}
\eta_{\text{M}} = \frac{P_S}{P_{\text{in}}}.
\label{eq:eff}
\end{equation}
To evaluate the performance of different pixel crosstalk models, we employ a large top-hat target light potential that requires SLM diffraction angles to approach the maximum steering angle of the SLM.
We define our target light potential in the image plane in units of the Fourier pixel pitch, $\SI{}{p_{\!\mathcal{F}}}$, in the computational image plane assuming twofold zero-padding of the electric field at the SLM, given by
\begin{equation}
    \SI{}{p_{\!\mathcal{F}}}=\frac{\lambda f}{2N\SI{}{p_{\text{SLM}}}} \approx \SI{6.54}{\micro\meter}
\end{equation}
after performing the FFT, where the zero-padded SLM plane contains $2N\times 2N$ pixels with pixel pitch, $\SI{}{p_{\text{SLM}}}$.
The maximum steering angle of the SLM corresponds to $\pm 1024\,\SI{}{p_{\!\mathcal{F}}}$ in the computational image plane.
The target light potential is a square with $600\,\SI{}{p_{\!\mathcal{F}}}$ side length with a dark border of $100\,\SI{}{p_{\!\mathcal{F}}}$ width, offset from the optical axis by $420\,\SI{}{p_{\!\mathcal{F}}}$ along the $x$- and $y$-direction (Fig.~\ref{subfig:square_potential}).
The target pattern is convolved with a Gaussian kernel of $2\,\SI{}{p_{\!\mathcal{F}}}$ width to remove sub-diffraction-limited edges in the target light potential.
As an initial phase guess for the CG algorithm, we use a linear phase to offset the potential from the optical axis by $480\,\SI{}{p_{\!\mathcal{F}}}$ in the x- and y-direction and a quadratic phase term with curvature $R=\SI{1.6}{mrad\per}{\SI{}{p_{\text{SLM}}}}^2$ \cite{Schroff2023}.

\section*{Backmatter}
\section{Funding}
Placeholder: We acknowledge support by the Engineering and Physical Sciences Research Council (EPSRC) through the Quantum Technology Hub in Quantum Computing and Simulation [grant number EP/T001062/1], the 2020-2021 Doctoral Training Partnership [EP/T517811/1],the Programme Grant DesOEQ [EP/P009565/1] and the New Investigator Grant [EP/T027789/1].

\section{Acknowledgments}
We would like to thank Jonathan Pritchard for helpful discussions.

\section{Disclosures}
The authors declare no conflicts of interest.

\section{Data Availability}
The datasets generated and analysed during the current study are not publicly available due to their large size (\SI{90}{GB} in images) but are available from the corresponding author upon reasonable request.
The Python code used in this study will be made available on GitHub at \url{https://github.com/paul-schroff/hologradpy}.

 \newpage

\bibliography{main}

\end{document}